\documentclass[10pt,oldtoc]{cernart}
\usepackage{times}
\usepackage{amsmath}
\usepackage{amsfonts}
\usepackage{amssymb}
\usepackage{pifont}
\usepackage{graphicx}
\usepackage{cite}
\usepackage{multicol}

\begin{document}
%=================================== symbols definition - start
\newcommand{\e}{\ensuremath{\mathrm{e}}}
\newcommand{\re}{\ensuremath{\mathrm{Re}}}
\newcommand{\im}{\ensuremath{\mathrm{Im}}}
\newcommand{\dm}{\ensuremath{\Delta m}}
\newcommand{\jpsi}{\ensuremath{J/\psi}}
\newcommand{\kn}{\ensuremath{K^0}}
\newcommand{\knb}{\ensuremath{\bar{K}{}^0}}
\newcommand{\bn}{\ensuremath{B^0}}
\newcommand{\bnb}{\ensuremath{\bar{B}{}^0}}
\newcommand{\T}{\ensuremath{\mathrm{T}}}
\newcommand{\U}{\ensuremath{\mathrm{U}}}
\newcommand{\ii}{\ensuremath{\mathrm{i}}}
\newcommand{\M}{\ensuremath{\mathrm{M}}}
\newcommand{\Laa}{\ensuremath{\mathrm{\Lambda^{11}}}}
\newcommand{\Lab}{\ensuremath{\mathrm{\Lambda^{12}}}}
\newcommand{\Lba}{\ensuremath{\mathrm{\Lambda^{21}}}}
\newcommand{\Lbb}{\ensuremath{\mathrm{\Lambda^{22}}}}
\renewcommand{\ket}[1]{\vert #1 \rangle}
\renewcommand{\bra}[1]{\langle #1 \vert}

%=================================== symbols definition - end
\thispagestyle{empty}
\begin{titlepage}
\docnum{CERN-OPEN-2013-042 v2}
\date{\today}
\vspace{20mm}
\title{{\large Observational Aspects of Symmetries of the Neutral B
        Meson System } }
\vspace{10mm}
\begin{center}
Maria Fidecaro$^1$,
Hans-J\"{u}rg Gerber$^2$,
Thomas Ruf$^1$ \\[10mm]
$^1$CERN, CH-1211 Geneva 23, Switzerland\\
$^2$ETHZ, IPP, CH-8093 Z\"urich, Switzerland \\[20mm]
\end{center}

\begin{abstract}
We revisit various results, which have been obtained by the BABAR and Belle Collaborations over the last thirteen years, concerning symmetry properties of the Hamiltonian, which governs the time evolution and the decay of neutral $B$ mesons. We find that those measurements, which established $CP$ violation in $B$ meson decay, 13 years ago, had as well established $T$ (time-reversal) symmetry violation. They also confirmed $CPT$  symmetry in the decay $(\T_{CPT}=0)$ and symmetry with respect to time-reversal $(\epsilon = 0)$ and to $CPT$  $(\delta = 0)$ in the \bn\bnb\ oscillation. Motion-reversal symmetry vs. time-reversal symmetry is discussed.
\end{abstract}
\end{titlepage}
\clearpage
%% main text %%%%%%%%%%%%%%%%%%%%%%%%%%%%%%%%% SECTION 1
\section{Introduction} \label{intro}
\small

A system of neutral mesons such as \bn , \bnb\ or \kn , \knb\ is a privileged laboratory for the study of weak-interaction's symmetries. Even though the phenomenological framework is well understood since long time
\cite{LOY,EL,BS,Kabir}, recent discussions in the physics community
\cite{SCHU}
show that it may be useful to revisit a few points, in order to fully (and correctly) exploit the experimental results. This process is then at the origin of the present note.

We focus on the \bn \bnb\ system, and refer to experimental results
\cite{Ba2001, Be2001, Be2002, Ba2002,Ba2012}
%%%%%%%%%%%%%%%%%%%%%%%5%%%%%%%%%% Be2002
that have been achieved by measurements of the decay products of
\bn\bnb\
pairs created in the entangled antisymmetric state
\begin{equation}\label{eq-1}
\ket{\Psi} = \left(\ket{\bn}\ket{\bnb} -\ket{\bnb}\ket{\bn}\right)/\sqrt2
\end{equation}
where the first $B$ in this notation moves in direction $\vec{p}$ and the second in direction $-\vec{p}$.

Within the Weisskopf-Wigner approximation \cite{LOY} the time evolution of a neutral $B$-meson, and its decay into a state $f$ is described by the amplitude $A_{Bf}$,
\begin{equation}\label{eq-2}
A_{Bf} = \ \bra{f}T \ \e^{-i\Lambda t}\ket{B}
\end{equation}
where \T\ and $\Lambda $ are represented by constant, complex $2\times2$ matrices $\T = (\T^{ij}) =  \bra{f^i}\T \ket{ B^j}$ \
and \ $\Lambda = (\Lambda^{ij}) =$  \hbox{$\bra{B^i} \Lambda \ket{B^j}$}, $i, j = 1 (2)$.
We consider experiments with final states $f^i= J/\psi K^i$ or $f^i= \mu^i\nu_{\mu} (\bar{\nu}_{\mu})X $.
Here $B^{1 (2)}$, $K^{1 (2)}$ and $\mu^{1 (2)}$ stand for the flavour eigenstates \bn~(\bnb), \kn~(\knb) and $\mu^+ (\mu^-)$ or $\e^+ (\e^-)$, respectively.

We recall that a symmetry is a property of the hermitian Hamiltonian ($H=H_0+H_{weak}$) of the Schr\"odinger equation which is  defined in a space sufficiently complete to include all the particle states under consideration, also the decay products \cite{LOY}. Thus the aim of the experiments is to establish properties of the weak interaction Hamiltonian $H_{weak}$ by measuring observable combinations of the elements of $\Lambda$ and of \T, which represent these properties.

As $CP$ violation implies $T$ and/or $CPT$ violation, we specifically consider the classical aim posed by the discoverers of $T$ violation \cite{SCHU2} "to express quantitatively the fraction of the observed $CP$ violation due to $T$ violation and $CPT$ violation separately".

In passing, we show that a more recent treatment, which attempts to define $T$-symmetry violation as "motion-reversal-symmetry violation", without reference to the weak interaction Hamiltonian \cite{Bebe}, is a special case within our phenomenology.\\
%
%%%%%%%%%%%%%%%%%%%%%%%%% SECTION 2
%
\section{Observables of Symmetries }\label{obser} %%%%%%%%%%%%%%%%%%%%%%%%%%%%%% TABLE 1 (START)
Together with a parametrization of the matrices $\Lambda$ and \T , the equations (1) and (2) are a sufficient basis for the description of the symmetry properties of the experimental results
\cite{Ba2001,Be2001,Be2002,Ba2002,Ba2012}. Symmetry properties of the Hamiltonian often manifest themselves in an especially simple and direct way in relations between measured quantities.
Here, {\it Table} 1 gives a summary, with  definitions and derivations as found in \cite{LOY, EL, BS, Kabir}, and the phase conventions of
\cite{EL}. Our approach is analogous to \cite{FG}. \\
\begin{center}
Table 1: \ {\it A symmetry of $H_{weak}$ implies vanishing values among the observables} $\Lambda_T,\ \Lambda_{CPT}, \ \T_T, \ \T_{CPT}$. \\ Channels are assumed to have one single amplitude. \\
%
%\begin{center}
\begin{tabular}{lll} \hline \\
Symmetry of $H_{weak}$ \ \ \ & $requires$ \ \ for\ \ the matrix \ $\Lambda$  & $requires$ \ \ for \ the matrix \ $T$ \\\\ \cline{1-3}\\
$T$ & $\Lambda_T \hspace{4.4mm}\equiv \ \mid \Lambda^{21}\mid ^2- \mid \Lambda^{12}\mid ^2 \ = 0 $ \ \ & $T_T \hspace{4.3mm}\equiv  Im(T^{11\star}\ T^{22}) \hspace{5.5mm} = 0$ \\
$CPT$ & $\Lambda_{CPT} \equiv \ \ \Lambda^{22} - \Lambda^{11} \hspace{9.3mm}= 0  $ & $T_{CPT} \equiv \ \mid T^{11}\mid ^2- \mid T^{22}\mid ^2 \ \ = 0 $ \\
$CP$ & $\Lambda_T \ \ \ \ \ = \ 0 \ \ \ \ and\ \ \ \Lambda_{CPT} \ \ \ \ = 0$ & $T_T \ \ \ \ \ = \ 0 \ \ \ \ and\ \ \ \ \ T_{CPT} \ \ \ = 0$\\\\ \cline{1-3}
\end{tabular}\\[5mm]
\end{center}
%%%%%%%%%%%%%%%%%%%%%%%%%%%%%% TABLE 1 END
%
Let us pose
\begin{eqnarray} %\label{eq3to4}
\Laa \ \ \ = \ \ \ m -\ii \gamma /2 - \delta \ \dm \ , \ \ \ \ \Lbb  &=& m - \ii \gamma /2+ \delta\ \dm \  , \\
\Lab \ \ \ = \ \ \ (1-2\epsilon) \ \dm/2 \ , \ \ \ \ \ \ \ \ \ \Lba  &=& (1 + 2 \epsilon) \ \dm/2 \ \ \ \
\end{eqnarray}
with real $m$, $\gamma$, \dm , $\epsilon$, and complex $\delta$.
For the observables of the symmetry violations in the matrix $\Lambda$, i.e. in the \bn\bnb\ oscillation, we deduce from eqs. (3), (4), and {\it Table} 1
\begin{eqnarray}
\label{eq5to6} \Lambda_T &=& 2 \ \epsilon  \ (\dm)^2  +  \mathcal{O}(\epsilon^2),      \\ \Lambda_{CPT} &=& 2 \ \delta \ \dm\ \ .
\end{eqnarray}

We note that, with eqs.(3), (4)  and (5), the difference of the widths of the eigenstates of $\Lambda $  becomes \\
$\Delta \Gamma = 2\dm \cdot \im (\sqrt{1-4\epsilon^2 + 4\delta^2} )$. This lets us recognize that, if $\Delta \Gamma = 0$, our matrix $\Lambda$  still allows
for a finite  $\epsilon$\hspace{0.1cm} ($|\epsilon| < 1/2$),
in accordance with
\cite{Ge2004}.  This is  in contrast to widely repeated affirmations \cite{Wo}, that $\Delta \Gamma = 0 $ would imply time-reversal symmetry of $\Lambda $, i.e. $\epsilon = \Lambda_T = 0$.

In terms of
$\Lambda =  \M - \frac{\ii}{2}\Gamma $ \ \ ($\M = \M^{\dagger}$ , \ $\Gamma = \Gamma^{\dagger}$),
$ \Lab  =   \ \mid \M^{12} \mid \e^{\ii \phi_\M} - \frac{\ii}{2 }\mid \Gamma^{12} \mid \e^{\ii \phi_{\Gamma}}$ ,
the relation to eqs. (3) to (5) is given by
$\epsilon =  -\frac{1}{4}\mid \Gamma^{12} \mid/ \mid \M^{12} \mid \times\sin(\phi_{\Gamma})$,
$\dm = 2 \mid \M^{12} \mid , \ \phi_\M = 0$ \  and $\ \Delta\Gamma \ \approx - 2 \mid \Gamma^{12}\mid  \cos(\phi_{\Gamma})$. We admit $\mid \Gamma^{12} \mid \ll \mid \M^{12} \mid $.

In order to calculate the amplitude $A_{Bf}$ in eq.(2), we need to evaluate the exponential in terms of $\Lambda$.  We do this by summing up the power series (as explained in \cite{FG}). Let \ $\U = (\U^{ij})=\e^{-i\Lambda t}$ \ and find
%
%%%%%%%%%%%%%%%%%%%%%%%%%%%%%% eqs7to9 (START)
%\begin{subequations}
\begin{eqnarray} \label{eq7to9}
          \U^{11} &= &\U_0 ( \cos(\omega t)+\ii \ 2\delta \sin(\omega t) \ ), \hspace{1cm}  \U^{22} = \U_0 ( \cos(\omega t)-\ii \ 2\delta \sin(\omega t) \ ),  \\
          \U^{12} &= &\U_0 (-\ii \ (1-2\epsilon) \sin(\omega t) ), \hspace{14.5mm} \U^{21}           = \U_0 (-\ii \ (1+2\epsilon) \sin(\omega t) ), \\
\mid \U_0 \mid ^2 &= & \ \e^{-\gamma t}, \nonumber      \\                                  \omega &= &\dm /2 + \mathcal{O}(\ \mid \delta \mid ^2, \epsilon ^2 \ ).
\end{eqnarray}
%\end{subequations}
%%%%%%%%%%%%%%%%%%%%%%%%%%%%%%%%%%%%%%%%%%%%%%%
For the matrix $(\T^{ij}) = \ (\bra{J/\psi K^i}\T \ket{B^j})$, we assume
\begin{equation} %label{eq10}
\T^{12} =\ \T^{21} = \ 0,
\end{equation}
with complex $\T^{11}, \T^{22}$, corresponding to the $"\Delta b = \Delta S$ rule". From {\it Table} 1\ , and with the (arbitrary) normalization \\
$ \mid \T^{11}\mid^2 + \mid \T^{22}\mid^2 \ = \ 2$\ , we deduce the useful identity among the (diagonal) elements of \T ,
\begin{eqnarray} %\label{eq11}
\T_T^2 + \T_{CPT}^2/4 + ( \re(\T^{11*}\T^{22}) \ )^2 \ \equiv (\mid \T^{11}\mid^2 + \mid \T^{22}\mid^2)^2/4  =  1 .
\end{eqnarray}
Results based on eqs. \ (1) \ to \ (11) will turn out to be sensitive to all the four symmetry parameters in {\it Table} 1.

Throughout this work, we assume that channels have one single amplitude. Two interfering amplitudes may fake non-vanishing values of $\T_{CPT}$ or $\T_T$, depending on their weak and strong phases, without the presence of the corresponding symmetry violations in the Hamiltonian.
%%%%%%%%%%%%%%%%%%%%%%%% SECTION 3
\section{Experiments}\label{sec:experimemt}
%%%%%%%%%%%%%%%%%%%%%%%%%%%%%%%%%%%%%%%%%%% SECTION 3.1
\subsection{General description}

Call $\mathcal{A}_{f_1 , f_2}(t) $ the amplitude for the decay of an entangled, antisymmetric \bn\bnb\ pair into a final state with the two observed particles $f_1 $ (at time $t_0$) and $f_2 $ (at later time $t> 0$).
With specific  choices  of the two final states \ $f_1, \ f_2$ , we can uniquely represent the complete set of results of the $CP$-, $T$-  and $CPT$-symmetry violation studies listed in {\it Table} 2 and performed by \cite{Ba2001,Be2001,Be2002} through \cite{Ba2012}, by making use of eq. (12) below \cite{EL,Day}, whose derivation we sketch here.
We note with (\cite{FG}, section 2.7), that the time evolution  acts on the two-particle state $\ket{\Psi}$ of eq. (1) solely by a multiplicative factor, which is independent of the symmetry violations under consideration, and which does not influence the decay properties of $\ket{\Psi}$. We may thus, without loss of generality, arbitrarily choose $t_0 = 0,\ t > 0$, and apply eq. (2) to the single-particle components in $\ket{\Psi}$, to obtain
\begin{eqnarray}
\mathcal{A}_{f_1 , f_2}(t) &=& \ \bra{f_1, f_2}\Psi\rangle  \\
                           &=& (\ \bra{f_1} \T \ket{B^0} \bra{f_2} \T \ e^{-i\Lambda t}\ket{\bar{B}^0}  - \notag \\
& & \phantom{(}\ \bra{f_1} \T \ket{\bar{B}^0} \bra{f_2} \T\ e^{-i\Lambda t}\ket{B^0}\ )/\sqrt2 \ . \notag
\end{eqnarray}
In rewriting (12), we can explicitly derive the formula for the state $\ket{S_{f_1} } $, which survived the decay to $f_1$, and its (single particle) time evolution and decay to $f_2$ as
\begin{eqnarray}
\mathcal{A}_{f_1 , f_2}(t) &\equiv& \mathcal{A}_{S_{f_1}, f_2} \notag \\
&=& \ \bra{f_2} \T\ e^{-i\Lambda t} \ket{S_{f_1}}
\end{eqnarray}
with
\begin{eqnarray} \ket{S_{f_1}}
&=& \ b \ \ket{B^0} + \ \ \bar{b} \ket{\bar{B}^0}  \\
b &=& - \bra{f_1} \T \ket{\bar{B}^0}/\sqrt2 \notag \\
\bar{b} &=& \phantom{-} \bra{f_1} \T \ket{B^0}/\sqrt2 \ . \notag
\end{eqnarray}

The variety of expected frequency distributions \hbox{$\mid \mathcal{A}_{f_1 , f_2}(t)\mid^2$} is displayed in {\it Table} 2.
We find that the parameters of the data analysis are the $T$ and $CPT$ violation parameters of the T matrix, $\T_T$ and
$\T_{CPT}$ , concerning the decay, and those, $p_i$, $q_i$, ($i=1,\, 2,\, 5,\, 6$), concerning mainly the
\bn\bnb oscillation matrix $\Lambda$. In the limit of $CP$ symmetry of $\Lambda$ the $p_i$, $q_i$ all vanish.
Then, $\T_T$ and $\T_{CPT}$\ are exactly associated each with its own proper time dependence: \ \ $\T_T$ with
$\pm\sin(\dm t)$, \ \ and \\
$\T_{CPT}$ with $\pm \cos(\dm t)$.

{\it Table} 2 also allows one to read off the relations of the measured distributions to the symmetry violating parameters of $\Lambda$ and T, as demonstrated below, and also to construct combinations of data which are true signatures for specific violations.
% %%%%%%%%%%%%%%%%%%%%%%%%%%%%%%%%%%%%%%%  SECTION 3.2
\subsection{The earlier results}
The experiments \cite{Ba2001, Be2001,Be2002,Ba2002} have measured in 2001/2 all the data sets listed in {\it Table} 2, and thereby discovered $CP$ violation in the matrix T. We show now that these data furthermore establish time-reversal symmetry violation in $H_{weak}$, and are compatible as well with $CPT$ symmetry of the T matrix as with $\epsilon = 0$, $\delta = 0$, i.e. $CP$ symmetry of $\Lambda$.

To this purpose we consult {\it Table} 2 and calculate
\begin{equation}
\{1\}-\{2\} = (p_1 - p_2) +  (\T_{CPT} - (p_1 -p_2))\cos(\dm t) + (2\T_T  \ +  \ (q_1-q_2))\sin(\dm t) . \notag
\end{equation}
Similarly, we calculate \{5\}--\{6\} and summarize the results as follows.
\begin{eqnarray}
CP_{S(L)} &\equiv &\ \mid \mathcal{A}_{\mu^-,J/\Psi K_{S(L)}^0}(t)\mid ^2 - \mid \mathcal{A}_{\mu^+,J/\Psi K_{S(L)}^0}(t)\mid ^2 \ \notag \\
&\propto & \hspace{15mm} 4 \epsilon \ \ \mp \ \ 4 Re(\delta) \ Re(T^{11*}T^{22}) \\ \notag
&+& \{ T_{CPT} \ \ - 4 \epsilon \ \ \pm \ \ 4 Re(\delta) \ Re(T^{11*}T^{22}) \} \cos(\Delta m\ t) \notag \\
&+& \hspace{27.5mm}      \{\pm 2 \ T_T \ -4 \ Im(\delta) \} \sin(\Delta m\ t)\ .
\end{eqnarray}
The experimental results for $CP_S$ and $CP_L$ show no time independent terms,
$ 4 \epsilon \mp 4 \re(\delta) \cdot \re(\T^{11*}\T^{22}) \approx 0 $, and no
$\cos(\dm\ t)$ signals, $\{ \T_{CPT} \ \ - 4 \epsilon \ \ \pm \ \ 4 \re(\delta)
\cdot \re ( \T^{11*} \T^{22}) \} \approx 0$. From this we conclude
$\epsilon \approx 0$, $4 \re(\delta) \cdot \re(\T^{11*}\T^{22}) \approx 0$, and
$\T_{CPT} \approx 0$. The $\sin(\Delta m\ t)$ amplitudes are equal but with opposite signs, and, in absolute value,
$< 2$,
implying $\im(\delta) \approx 0$ and $\mid \T_T\mid ^2 < 1 $.
From (11) now follows \ $\re(\T^{11*}\T^{22}) \ne 0$ \ and thus $\re(\delta) \approx 0$.
The $p_i$ and $q_i$ defined in {\it Table} 2 are thus all compatible with  zero.

Quantitative results for $\T_T$ and $\T_{CPT}$ may be read off from
\cite{Ba2001} and
\cite{Be2001,Be2002}, who analyze their data also with two free parameters
\cite{Ba2001,Be2002}, corresponding to $\T_T$ and $\T_{CPT}$.
%
%
%%%%%%%%%%%%%%%%%%%%%%%%%%%%%%%%%%%%% TABLE 2 (START)
\begin{table*} Table 2: {\it The measurements, classified according to eq. (12)}. \\
General expressions for the expected frequency distributions in terms of $\T_{CPT}, \T_T, \epsilon , \delta .$
In the limit $\epsilon = \delta = 0$, they are all of the form
$( \ \ 1 \ \pm \frac{1}{2} \T_{CPT} \ \cos(\Delta m\ t)\ \pm \T_T \ \sin(\Delta m\ t) \ \ ) \ e^{-\gamma t}.$
$\mu^-$ is a shorthand for $\mu^- \bar{\nu}_\mu X$ or \ $e^- \bar{\nu}_e X$, \
$\mu^+$ \ for \ $\mu^+ \nu_\mu X$, etc.
By the "$\Delta b = \Delta Q$ rule", a $B^0$($\bar{B^0}$) decays semileptonically always into $\mu^+ + ...$ \ ($\mu^- + ...$).
$\mid K_{S(L)}> \ = (\mid K^0> \pm \bar{\mid K^0}>)/\sqrt2 $ \ has been used.
All 10 measurements have been performed.
%
%%%%%%%%%%%%%%%%%%%%%%%%%%%%%%%%%%%
\vspace{3mm}
\begin{tabular}{llll} \hline \\
Name of measurement & $1^{st}$ decay & $2^{nd}$ decay & $\mid
\mathcal{A}_{f_1 , f_2}(t)\mid^2  \ \propto \ a + b\ \cos(\Delta m\ t) + c\ \sin(\Delta m\ t)             $   \\
\empty & $f_1$ & $f_2$ & $\phantom{aa}a$ \hspace{18.5 mm} \phantom{aaaaa}$b$ \hspace{20 mm} $c$ \\
\cline{1-4}\\
$B^0 \rightarrow  K_S^0$ \ \ \ \ \ \{1\} & $\mu^-$  \ \ & $J/\psi K_S^0$ & $1 + p_1$ \hspace{9mm} $ + \ \ {\textstyle\frac{1}{2}}\ \T_{CPT} - p_1 \hspace{9 mm} + \T_T +q_1 \ $ \\
$\bar{B}^0 \rightarrow  K_S^0$ \ \ \ \ \ \{2\} & $\mu^+$ & $J/\psi K_S^0$ & $1 + p_2$ \hspace{9mm} $ - \ \ {\textstyle\frac{1}{2}}\ \T_{CPT} - p_2 \hspace{9 mm} - \T_T +q_2 \ $  \\
$K_L^0 \rightarrow  \bar{B}^0$ \ \ \ \ \ \{3\} &  $J/\psi K_S^0$ & $\mu^- $ & $1 + p_1$ \hspace{9mm} $ + \ \ {\textstyle\frac{1}{2}}\ \T_{CPT} - p_1 \hspace{9 mm} - \T_T -q_1 \ $ \\
$K_L^0 \rightarrow  B^0$ \ \ \ \ \ \{4\} &  $J/\psi K_S^0$ & $\mu^+ $ & $1 + p_2$ \hspace{9mm} $ - \ \ {\textstyle\frac{1}{2}}\ \T_{CPT} - p_2 \hspace{9 mm} + \T_T -q_2 \ $ \\ & & & \\ \cline{1-4}\\
$B^0 \rightarrow  K_L^0$ \ \ \ \ \ \{5\} & $\mu^-$ & $J/\psi K_L^0$ & $1 + p_5$ \hspace{9mm} $ + \ \ {\textstyle\frac{1}{2}}\ \T_{CPT} - p_5 \hspace{9 mm} - \T_T +q_5 \ $  \\
$\bar{B}^0 \rightarrow  K_L^0$ \ \ \ \ \ \{6\} & $\mu^+$ & $J/\psi K_L^0$ & $1 + p_6$ \hspace{9mm} $ - \ \ {\textstyle\frac{1}{2}}\ \T_{CPT} - p_6 \hspace{9 mm} + \T_T +q_6 \ $\\
$K_S^0 \rightarrow  \bar{B}^0$ \ \ \ \ \ \{7\} &  $J/\psi K_L^0$ & $\mu^- $ & $1 + p_5$ \hspace{9mm} $ + \ \ {\textstyle\frac{1}{2}}\ \T_{CPT} - p_5 \hspace{9 mm} + \T_T -q_5 \ $\\
$K_S^0 \rightarrow   B^0$ \ \ \ \ \ \{8\} &  $J/\psi K_L^0$ & $\mu^+ $ & $1 + p_6$ \hspace{9mm} $ - \ \ {\textstyle\frac{1}{2}}\ \T_{CPT} - p_6 \hspace{9 mm} - \T_T -q_6 \ $ \\
$\bnb\ \rightarrow \bn\ $ \ \ \ \  \{9\} & $\mu^+$        & $\mu^+ $ & $\frac{1}{2} \ (1-4\epsilon)$\hspace{4mm}  $- \ \ \frac{1}{2} \ (1-4\epsilon) $ $\hspace{15 mm} 0 $ \\
$ \bn\ \rightarrow \bnb\ $  \ \ \  \{10\} & $\mu^-$       & $\mu^- $ & $ \frac{1}{2} \ (1+4\epsilon)$\hspace{4mm} $- \ \ \frac{1}{2} \ (1+4\epsilon) $ $\hspace{15 mm} 0$ \\
& & & \\
\cline{1-4}
\end{tabular}

\begin{tabular}{ll}
The terms with \ $\epsilon $ \ and $ \delta $  \ \ \ \ \ \ (upper signs for $p_1, p_5, q_1, q_5 $ ).  & \hspace{48mm} \empty    \\\\
$p_1 (p_2) = \epsilon \ (\pm 2-\T_{CPT}) \mp 2 \re(\delta)\cdot \re(\T^{11*} \T^{22})- 2 \im(\delta) \T_T $             \\$p_5 (p_6) = \epsilon \ (\pm 2-\T_{CPT}) \pm 2 \re(\delta) \cdot \re(\T^{11*} \T^{22})+ 2 \im(\delta) \T_T $  \\
$q_1 (q_2) = \ \ \ \epsilon \cdot 2\ \T_T \ -\im(\delta) (\pm 2+\T_{CPT}) $
&  Identity:                 \\
$q_5 (q_6) =  -\epsilon \cdot 2\ \T_T \ -\im(\delta) (\pm 2+ \T_{CPT})     $
&  $q_1 + q_6 - (q_2 + q_5) = 0  $ \\
&  \\
\cline{1-2}\\
\end{tabular}
\end{table*} %%%%%%%%%%%%%%%%%%%%%%%%%%%%%%% TABLE 2 (END)
%%%%%%%%%%%%%%%%%%%%%%%%%%%%%%%%%% TABLE 3 (START)
\begin{table*}
Table 3: {\it A selection of expectations for the experiment of Ref.
\cite{Ba2012}}. \\
Due to the presence of $\T_{CPT}$, of the $p_i$ and  $q_i$, our results
contradict the attempt
\cite{Bebe,Bebe2} to define the differences $\{2a\}$ to  $\{2d\}$, each as a
signature for  $T$ violation. In the lower part, signatures for $T$- and $CPT$- symmetry violations are indicated.

%%%%%%%%%%%%%%%%%%%%%%%%%%%%%%%%%%%
\begin{tabular}{llll} \hline \\
Display in \cite{Ba2012} & Rates compared \hspace{10mm} & Expected \ \ $\propto
$ &
$ a  + b  \cos(\Delta m\ t) + c \sin(\Delta m\ t)$  \\
\empty              & \empty                         & \hspace{5mm}$ a $ &
\hspace{15mm} $ b $  \hspace{30mm} $ c $ \\
\cline{1-4}\\
{\it Figure} \ \   2a   & $\{2\} - \{7\} \equiv \{2a\} $ & $p_2 -p_5$ & $ - \ \T_{CPT} - (p_2 - p_5)$  \hspace{10mm} $ - 2 \ \T_T \ + q_2 + q_5\   $ \\
\phantom{{\it Figure}\ \ }  2b   & $\{4\} - \{5\} \equiv \{2b\} $ & $p_2 -p_5$ & $ - \ \T_{CPT} - (p_2 - p_5)$   \hspace{10mm} $ + 2\ \T_T \ - q_2 - q_5\   $   \\
\phantom{{\it Figure}\ \ }  2c  & $\{6\}-\{3\} \equiv \{2c\}    $ & $p_6 -p_1$ & $ - \ \T_{CPT} - (p_6 - p_1)$   \hspace{10mm} $ + 2 \ \T_T \ + q_1 + q_6  $ \\
\phantom{{\it Figure}\ \ }  2d  & $\{8\}-\{1\} \equiv \{2d\}    $ & $p_6 -p_1$ &
$ - \ \T_{CPT} - (p_6 - p_1)$   \hspace{10mm} $ - 2 \ \T_T \ - q_1 - q_6 $
\\\\
\cline{1-4} \end{tabular} \\
\begin{tabular}{llll} & & &  \\
Signatures\ are & for \ $T_T$          & $\hspace{5mm} - 8\ \T_T
\sin(\Delta m\ t) $
& $\propto \ \{2a\}-\{2b\}-\{2c\}+\{2d\} $  \\
\empty         & \ \ \phantom{for}$\T_{CPT} $ & $\hspace{5mm} - 4\ \T_{CPT}  $
& $\propto \ \{2a\}+\{2b\}+\{2c\}+\{2d\} \quad (t=0) $ \phantom{tttt=========}\\
\empty         & \ \ \phantom{for}$\Lambda_{T} $
& $ \hspace{5mm}\phantom{+} 4\ \epsilon $
& $\approx (\{10\}-\{9\})\ / \ (\{10\}+\{9\})$ \\\\
\cline{1-4}
\end{tabular}
\end{table*} %%%%%%%%%%%%%%%%%%%%%%%%% TABLE 3 (END)

The experiment
%\cite{Ba2002,Ba2006}
\cite{Ba2002} has set a stringent limit on $T$-symmetry violation in the $\Lambda$  matrix of the \bn\bnb\ system with a direct measurement of $\epsilon$. See \it Table\rm\ 2 (entries \{9\} and \{10\}) and
\it Table\rm\ 3.
The method is analogous to the one of the CPLEAR experiment \cite{CPLEAR,BT} for the \kn\knb\ system, where also a signature for $T$-violation ("Kabir asymmetry") has been directly measured. The experiments make use of the general identity, valid in two dimensions (see \cite{FG}),
$  \Lambda^{21}/\Lambda^{12} \equiv (\e^{-i\Lambda t})^{21} / (\e^{-i\Lambda t})^{12} = \U^{21}/\U^{12} $ \ \ \
from which
\begin{equation}\label{eq. 14}
\epsilon\ \approx \frac{1}{4} \frac{\mid \Lambda^{21}\mid ^2 - \mid \Lambda^{12}\mid ^2}
                 {\mid \Lambda^{21}\mid ^2 + \mid \Lambda^{12}\mid ^2}
\equiv            \frac{1}{4} \frac{\mid \U^{21}\mid ^2 - \mid \U^{12}\mid ^2 }
                 {\mid \U^{21}\mid ^2 + \mid \U^{12}\mid ^2}  \notag \\
=                \frac{1}{4} \frac{\mid \mathcal{A}_{\mu^-\mu^-}\mid ^2 - \mid \mathcal{A}_{\mu^+\mu^+}\mid ^2}
                 {\mid \mathcal{A}_{\mu^-\mu^-}\mid ^2 + \mid \mathcal{A}_{\mu^+\mu^+}\mid ^2}\ ,
\end{equation}
the connection from the data to the $T$- symmetry violation signal $\epsilon$ ,
follows   - without any assumptions on $CPT$ symmetry or on the value of $\Delta \Gamma $ of the $\Lambda$ matrix.

A reanalysis of the results in 2007 of the BABAR and Belle collaborations by
\cite{Eze} has shown that the data contradict {\em motion-reversal symmetry}
(see \cite{SCHU}) in the \bn\bnb\  system.

In summary, the discovered $CP$ violation in the \bn\bnb\ system is $T$-symmetry violation in the decay-amplitude matrix \T , $\T_T \ne 0 $  with $\T_{CPT} \approx 0$.
%The matrix $\Lambda$ is confirmed to be $CP$ symmetric.
In the \kn\knb\ system, however, the $CP$-violation  is $T$-symmetry violation in oscillations, $\Lambda_T \ne 0$ with $\Lambda_{CPT} \approx 0$ .
%
 %%%%%%%%%%%%%%%%%%%%%%%%%%%%%%%%%%%% SECTION 3.3
\subsection{Recent results}
The analysis by \cite{Ba2012} is based on \cite{Bebe} with novel notions of $CPT$-, $CP$-, \ and $T$-symmetry, which, in contrast to the classical definitions \cite{LOY}, are not related to properties of the weak interaction Hamiltonian, but to comparisons of surviving states $\ket{S_{f_1}}$ with suitably motion-reversal transformed ones of type $\ket{S_{f'_1}}$. The novel definitions are less general than the classical ones as they need the assumption of $\T_{CPT} = 0$. This new analysis then becomes a special case of our present work, and in turn looses the possibility to address the "classical aim", mentioned in our Introduction. (Details below).

To prove that the phenomenology of \cite{Bebe} uses $\T_{CPT} = 0$, it is sufficient to express their
eq.(A.5 of \cite{Bebe}) in terms of the elements of the matrix $\T$, $\T^{11}$ and $\T^{22}$, to find \\
\centerline{$\alpha \beta^* = -1 = - \mid \T^{11} \mid ^2 / \mid \T^{22} \mid ^2 $ or \ \ $\T_{CPT} = 0$.}

The work of \cite{Bebe} specifies 3 sets of 4 pairs of measurements, whose comparisons are supposed to indicate the violations of the 3 symmetries mentioned above.
(See \it Tables \rm 1, 2, 3 of \cite{Bebe}). Each of the 24 measurements is completely determined by the products of the first and the second decay of the antisymmetric, entangled $B^0 \bar{B}^0$ pair. Their amplitudes are thus uniquely given by our eq. (12). The corresponding rates are listed in our \it Table\rm\ 2, labeled \{1\} to \{8\}.

The envisaged \ $T$-violating comparisons, labelled \{2a\} to \{2d\} in
\it Table\rm\ 3, depend also on $\T_{CPT}$, \ and thus contradict the affirmation in
\cite{Ba2012}, that "Any difference in these two rates is evidence for $T$-symmetry violation",
since a $T$-symmetric, $CPT$-violating Hamiltonian $H_{weak} \ $($\T_T = 0, \T_{CPT} \neq 0$) would just also create such rate differences.

The $CP$-violating comparisons in \it Table\rm\ 2 of \cite{Bebe} also depend on $\T_{CPT} \cos(\Delta mt)$ and on $\T_T \sin(\Delta mt)$.
This confirms that $T$- and/or $CPT$-violation imply $CP$-violation. $T$-violation in the (2 by 2 dimensional) $B^0 \bar{B}^0$ system is thus never independent of $CP$ violation.
See also \cite{Ge2004}.

The $CPT$-violating comparisons in \it Table\rm\ 3 of \cite{Bebe} neither depend on $\T_{CPT}$ nor on $\T_T$, and are thus, contrary to the authors' intentions, unable to detect $CPT$ symmetry violation in the matrix $\T$.

Nevertheless, the measured frequency distributions \{2a\} to \{2d\} show a dominant $\sin(\Delta m \ t)$ time-dependence, meaning, for this reason, that $\T_{CPT} \approx 0$, and with the previous knowledge about the vanishing of the $q_i $, that $\T_T \ne 0$, i. e. $T$-symmetry violation is confirmed. (More combinations are discussed in
\cite{SCHU}). In the lower part of {\it Table} 3, we indicate rate combinations which are true signatures of $T$- or $CPT$- symmetry violations.
%%%%%%%%%%%%%%%%%%%%%%%%% SECTION 4
\section{Conclusion}\label{sec:concl}

The experiments \cite{Ba2001} and \cite{Be2001} have discovered $CP$ violation in the \bn\bnb\ system. Our analysis shows that this $CP$ violation is dominantly $T$ violation, with the same statistical significance. Furthermore,  their data sets contain the information which allows for the estimation of all symmetry-violating parameters indicated in {\it Table} 1. $CP$ symmetry of the matrix $\Lambda$, which governs the $B^0\bar{B^0}$ oscillation, is confirmed.

The novel definitions of the symmetries ($CP,\ T, \ CPT $) used by
\cite{Bebe,Bebe2} are more restrictive than the classical ones \cite{LOY}.
%
%%%%%%%%%%%%%%%%%%%%%%%%%% SECTION 5
\section*{Acknowledgment}
We are grateful for numerous, very helpful discussions with K. R. Schubert during and around the MITP workshop in Mainz, April 2013.
\newpage
%
%\section*{References}

\end{document}